\documentclass[Journal,onecolumn]{IEEEtran}
\usepackage{amsmath}
\usepackage{theorem}
\usepackage{amsmath}
\usepackage{mathrsfs}
\usepackage{accents}
\usepackage{url}
\usepackage{tabularx}
\usepackage{dsfont}
\usepackage{float}
\usepackage{amsmath}
\usepackage{algorithm}
\usepackage{algorithm}
\usepackage{algorithmic}
\usepackage{epsfig}
\usepackage{amsfonts}
\usepackage{amssymb}
\usepackage{mathrsfs}
\usepackage{euscript}
\usepackage{graphicx}
\usepackage{multirow}
\usepackage{cite}
\usepackage{placeins}
\usepackage{caption}
\usepackage{color,soul}
\usepackage{soul}
\usepackage{subcaption}
\usepackage{lipsum} 
\usepackage{filecontents} 
\usepackage{amsmath}
\usepackage{theorem}
\usepackage{amsmath}
\usepackage{mathrsfs}
\usepackage{accents}
\usepackage{setspace}
\usepackage{url}
\usepackage{algorithm}
\usepackage{algorithmic}
\usepackage{epsfig}
\usepackage{amsfonts}
\usepackage{amssymb}
\usepackage{mathrsfs}
\usepackage{euscript}
\usepackage{graphicx}
\usepackage{multirow}
\usepackage{cite}
\usepackage{amsmath}
\usepackage{theorem}
\usepackage{amsmath}
\usepackage{mathrsfs}
\usepackage{accents}
\usepackage{url}
\usepackage{tabularx}
\usepackage{float}
\usepackage{amsmath}
\usepackage{algorithm}
\usepackage{algorithm}
\usepackage{algorithmic}
\usepackage{epsfig}
\usepackage{amsfonts}
\usepackage{amssymb}
\usepackage{mathrsfs}
\usepackage{euscript}
\usepackage{graphicx}
\usepackage{multirow}
\usepackage{cite}
\usepackage{placeins}
\usepackage{caption}
\usepackage{color,soul}
\usepackage{soul}
\usepackage{subcaption}
\usepackage{lipsum} 
\usepackage{filecontents} 
\newtheorem{theorem}{Theorem}
\makeatletter

\def\subsubsection{\@startsection{subsubsection}
                                 {3}
                                 {\z@}
                                 {0ex plus 0.1ex minus 0.1ex}
                                 {0ex}
                                 {\normalfont\normalsize\itshape}}
\makeatother

\usepackage[T1]{fontenc}
\usepackage[USenglish,UKenglish,french,spanish,italian]{babel}
\usepackage[nodayofweek,level]{datetime}
\newcommand{\mydate}{\formatdate{13}{12}{2017}}

\graphicspath{{./}{figures/}}

\ifCLASSINFOpdf
\else
\fi
\begin{document}

\begin{titlepage}

\begin{tabular}{l        r}

\includegraphics[bb=20bp 00bp 500bp 450bp,clip,scale=0.3]{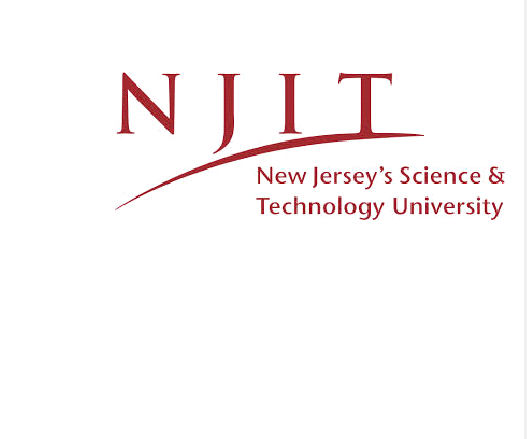} \hspace{6cm} & \includegraphics[bb=0bp -200bp 500bp 550bp,clip,scale=0.2]{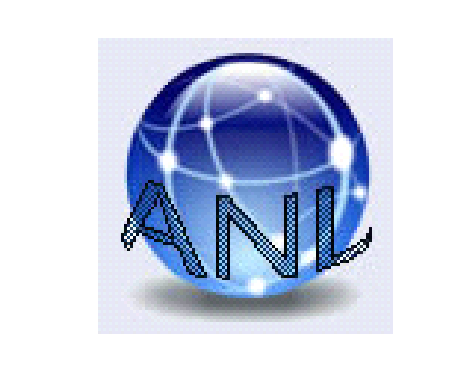}

\end{tabular}

\begin{center}

\textsc{\LARGE Edge Computing Aware NOMA for 5G Networks}\\[1.5cm]

{\Large \textsc{Abbas Kiani}}\\ 
{\Large \textsc{Nirwan Ansari}}\\
[2cm]

{}
{\textsc{TR-ANL-2017-007}\\
\selectlanguage{USenglish}
\large \mydate} \\[3cm]

{\textsc{Advanced Networking Laboratory}}\\
{\textsc{Department of Electrical and Computer Engineering}}\\
{\textsc{New Jersy Institute of Technology}}\\[1.5cm]
\vfill

\end{center}

\end{titlepage}

\selectlanguage{USenglish}

\begin{abstract}
With the fast development of Internet of things (IoT), the fifth generation (5G) wireless networks need to provide
massive connectivity of IoT devices and meet the demand for low latency.
To satisfy these requirements, Non-Orthogonal Multiple Access (NOMA) has
been recognized as a promising solution for 5G networks
to significantly improve the network capacity.
In parallel with the development of NOMA techniques, Mobile Edge Computing (MEC) is becoming one of the key emerging technologies to reduce the latency and improve the Quality of Service (QoS) for 5G networks.
In order to capture the potential gains of NOMA in the context of MEC, this paper proposes an edge computing aware NOMA technique which can enjoy the benefits of uplink NOMA in reducing MEC users' uplink energy consumption.
To this end, we formulate a NOMA based optimization framework which minimizes the energy consumption of MEC users via optimizing the user clustering, computing and communication resource allocation, and transmit powers. In particular, similar to frequency Resource Blocks (RBs), we divide the computing capacity available at the cloudlet to computing RBs. Accordingly, we explore the joint allocation of the frequency and computing RBs to the users that are assigned to different order indices within the NOMA clusters.
We also design an efficient heuristic algorithm for user clustering and RBs allocation, and formulate a convex optimization problem for the power control to be solved independently per NOMA cluster.
The performance of the proposed
NOMA scheme is evaluated via simulations.

\end{abstract}
\section{Introduction}\label{sec:Introduction}
With the fast development of mobile Internet and Internet of Things (IoT), mobile data traffic is anticipated to witness explosive growth in the years to come.
To support this unprecedented growth, both academic and industrial communities have conducted extensive research to design the fifth generation (5G) wireless networks. The 5G networks are to offer significant improvements of wireless network capacity and the user experience~\cite{thompson20145g}, and demand spectral-efficient multiple access techniques.
To this end, Non-Orthogonal Multiple Access (NOMA) techniques~\cite{saito2013non} have
been recognized as promising solutions for 5G and have attracted extensive research recently.
In contrast with Orthogonal Multiple Access (OMA)
techniques, where the radio resources are allocated orthogonally
to multiple users, NOMA allows multiple users to share the same resources. By serving multiple users simultaneously over the same radio resources, more users can be supported, thus leading to a significant increase in the network capacity. This improvement
is nevertheless available at the expense of intra-cell interference as well as additional complexity at the receiver side.
To deal with the intra-cell interferences and the complexity, NOMA splits the users in the power domain based on their respective channel conditions and employs efficient Multi-User
Detection (MUD) technique such as Successive Interference Cancellation at the receiver side~\cite{kim2013non}.

In parallel with the explosive growth of mobile data traffic, our daily life witnesses a significant increase in demands for running sophisticated applications in the mobile devices for social networking, business, etc.~\cite{barbarossa2014communicating}.
Moreover, in future user-centric 5G networks, the IoT users participate
in sensing and computing tasks, and computation-intensive tasks
need to be offloaded to either the cloud or the computing
resources at the edge. To this end, Mobile Edge Computing (MEC), which is being standardized by an Industry Specification Group (ISG) lunched by the European Telecommunications Standards Institute (ETSI)~\cite{hu2015mobile}, is recognized as one of the key emerging technologies for 5G networks. The idea of MEC is to provide computing capabilities in proximity of users and within the Radio Access Network (RAN), thereby reducing the latency and improving the Quality of Service (QoS)~\cite{hu2015mobile}.

In this paper, we focus on two aforementioned emerging technologies of 5G, i.e., NOMA and MEC, and propose a novel MEC aware NOMA technique for 5G networks. Our proposed scheme is motivated by the fact that the joint allocation of communication and computing resources greatly improves the performance of the system. In other words, it may happen that one type of resources is wasted due to congestion of other type of resources. While several works such as~\cite{di2013joint,yu2016joint} have investigated the joint allocation of computing and communication resources, non of the existing works consider a joint optimization technique in the context of NOMA with consideration of intra-cell interferences. To this end, the current study aims to address the aforementioned issue by proposing a joint optimization technique to allocate the computing and communication resources based on the requirements of both MEC and NOMA.

\textbf{Contributions:} We have made three major contributions. 1) We propose a novel NOMA augmented edge computing model that captures the gains of uplink NOMA in MEC users' energy consumption. Specifically, we design a NOMA based optimization framework that minimizes the energy consumption of MEC users via optimizing the user clustering, computing and communication resource allocation, and transmit powers. To this end, similar to frequency Resource Blocks (RBs), we define the notion of computing RBs and investigate the joint allocation of the frequency and computing RBs. More importantly, we consider a time constraint for each edge computing task, and accordingly the minimum data rate requirement of each user is established based on its deadline.
2) We design an efficient heuristic algorithm for user clustering and RBs allocation. Moreover, we formulate a convex optimization problem for the transmission power control to be solved independently per NOMA cluster.
3) We evaluate the performance of our proposed NOMA scheme and the heuristic algorithm via extensive simulations in which we show the benefits of NOMA in reducing the MEC users' uplink energy consumption. We also evaluate the effects of computing capacity and its division strategy of computing RBs on the total energy consumption.

\textbf{Related works:}
The related works to this paper include MEC, NOMA, and sub-carrier scheduling.
In the past few years, a large and cohesive body of work investigated the major challenges of MEC and
the researchers came up with a variety of policies and algorithms.
Recently, Chiang \textit{et al.}~\cite{chiang2016fog} summarized the opportunities and challenges of edge computing in the networking context of IoT and Gonzalez \textit{et al.}~\cite{gonzalez2016fog} explored the applications of edge computing in IoT.
Yu \textit{et al.}~\cite{yu2016joint} proposed a joint subcarrier and CPU time allocation algorithm for
MEC. A hierarchical MEC model designed based on the principle
of LTE-Advanced backhaul network is introduced in~\cite{kiani2017towards} in which the so called field, shallow, and deep cloudlets are located hierarchically in three different tiers of the network. A task scheduling scheme for code partitioning over time and the hierarchical cloudlets is also proposed in~\cite{kiani2017optimal}.
Moreover, a novel approach to mobile edge
computing for the IoT architecture is presented in~\cite{sun2}.
A hybrid architecture that harnesses the synergies between edge caching and C-RAN is proposed in~\cite{tandon2016harnessing}.

Lee \textit{et al.}~\cite{lee2009proportional} explored the fundamental problem of LTE
Single-carrier FDMA uplink scheduling by adopting the conventional time domain
proportional fair algorithm.
As discussed earlier, this paper proposes NOMA based model for MEC. Recently, several research studies have identified the potential benefits of NOMA in both the downlink and uplink. For instance, Al-Imari~\textit{et al.}~\cite{al2014uplink} proposed a NOMA scheme for uplink that allows more than one user to share the same subcarrier while a joint processing is implemented at the receiver to detect the
users' signals. Zhang~\textit{et al.}~\cite{zhang2016uplink} designed
an uplink power control scheme where eNB distinguishes the multiplexing users in
the power domain, and theoretically analyzed the outage performance and the achievable sum-rate of the proposed scheme.
Liang~\textit{et al.}~\cite{liang2017non} proposed the so called non-orthogonal random access (NORA)
scheme based on SIC to tackle the the access congestion problem. In NORA, the difference of time of arrival is used to
identify multiple users. Shipon~\textit{et al.}~\cite{ali2016dynamic} also designed a sum-throughput maximization problem under transmission power constraints for both uplink and downlink NOMA. Moreover, Tabassum~\textit{et al.}~\cite{tabassum2017modeling}
characterized the rate coverage probability of a user in
NOMA cluster with a given rank as well as the mean rate coverage probability of all users in the cluster
for perfect SIC, imperfect SIC, and imperfect worst case SIC.

While there are numerous research activities that investigate NOMA technique and its benefits in 5G networks, there is no prior work that study the advantages of NOMA in the context of edge computing. To this end, the current studied aims at proposing a novel edge computing aware NOMA model to reduce the uplink energy consumption of MEC users via utilizing the gains of uplink NOMA. Moreover, we take into consideration of the deadline requirements of MEC users in the user clustering.

The rest of the paper is organized as follows. Section~\ref{sec:System Model}
describes the system model and problem formulation. We propose our optimization framework and the corresponding heuristic algorithms in Section~\ref{sec:optimization}. Finally, Sections~\ref{sim:results} and~\ref{conclude} present numerical results and conclude the paper, respectively.

\section{System Model and Problem Formulation}\label{sec:System Model}
We consider a single-cell scenario, where one eNB equipped with a cloudlet serves the uniformly distributed edge computing users.
Denote $\mathcal{U}=\{1,...,U\}$ as the set of users each with a task to be offloaded to the cloudlet via eNB.
In the following, the users and tasks are used interchangeably.
Each task $u$ is characterized by the workload $\lambda_{u}$, i.e., the number of CPU cycles required to complete the execution of the task, and the input $L_u$, i.e., the number of bits that must be transferred from the user to the eNB.

\subsection{Communication Resources}
We assume that the available bandwidth is divided into a set of
frequency resource blocks $\mathcal{R}_f=\{1,...,M_f\}$ and the bandwidth of each resource block is $B$.
According to the NOMA schemes, it is assumed that the users transmit over the resource blocks in an non-orthogonal manner, i.e., more than one user can share the same resource blocks. Therefore, the users are assumed to be divided into different groups called the NOMA clusters, where a set of frequency RBs are allocated to each NOMA cluster.
Denote $\mathcal{I}=\{1,...,N\}$ as the set of NOMA clusters and $\beta^{r,i}$ as the binary variable to indicate the allocation of RB $r\in\mathcal{R}_f$ to NOMA cluster $i\in\mathcal{I}$. Here, $\beta^{r,i}=1$ if RB $r$ is allocated to NOMA cluster $i$, and $\beta^{r,i}=0$ otherwise.
Given the principles of NOMA in which at least two users must share the same frequency resource blocks, we set $N\leq\lfloor\frac{U}{2}\rfloor$.

We assume that an efficient MUD technique such as SIC~\cite{kim2013non} is applied at the eNB to decode the message signals in which the users are required to be ordered in each NOMA cluster. We define $\mathcal{J}=\{1,2,...,u_{max}\}$ as the set of the orders in a cluster. Here, $u_{max}$ is defined as the maximum number of users allowed to share a RB, hereby, reducing the complexity at the receiver side. It is also assumed that $Nu_{max}\geq U$.
According to the principles of MUD techniques~\cite{al2014uplink}, the message of the $j$-th user in the cluster is decoded before all the users with higher indices.
Consequently, the $j$-th user of a cluster experiences interference from all the users $\{j+1,j+2,...\}$ in that cluster.
In other words, the first user to be decoded ($j = 1$) will see interference
from all the other users $j = 2,...,u_{max}$, and the second user to
be decoded will see interference from the users $j = 3,...,u_{max}$,
and so on.

Denote $p_u^r$ as the transmission power of user $u$ over RB $r$ and $\alpha_u^{i,j}$ as the binary variable to indicate the assignment of user $u$ to the $j$-th order in cluster $i$. Here, $\alpha_u^{i,j}=1$ if the assignment occurs, and $\alpha_u^{i,j}=0$ otherwise.
The achievable data rate of user $u$ is given by
\begin{equation}\label{equ1}
R_u=\sum_{i\in\mathcal{I}}\sum_{j\in\mathcal{J}}\alpha_u^{i,j}\sum_{r\in\mathcal{R}_f}\beta^{r,i}B\log(1+\frac{h_u^rp_u^r}{\sigma^2+\sum_{k\in\mathcal{U}\setminus u}\sum_{l=j+1}^{u_{max}}\alpha_k^{i,l}h_k^rp_k^r})
\end{equation}
where $h_u^r$ denotes the channel gain
between user $u$ and the eNB on RB $r$, and $\sigma^2$ is the noise power. Note that by $h_u^r$, we assume
that the channel conditions vary across RBs as well as users.
Therefore, the uplink transmission time of task $u$ is
\begin{eqnarray}\label{equ2}
T_u=\frac{L_u}{R_u}
\end{eqnarray}
and the energy consumption of user $u$ is given by
\begin{eqnarray}\label{equ3}
E_u=T_u\sum_{r\in\mathcal{R}_f}p_u^r
\end{eqnarray}

\subsection{Computing Resources}
Analogous to the communication resources, we assume the computing capacity of the cloudlet is divided into different computing RBs. For example, one computing RB can be one virtual machine or one core. Denote $\mathcal{R}_c=\{1,...,M_c\}$ as the set of all computing RBs. We also assume the capacity of one computing RB is equal to $C$ CPU cycles per second.
It is assumed that a number of computing RBs is allocated to each order index of each NOMA cluster.
Therefore,
the computing time of task $u$ is
\begin{eqnarray}\label{equ4}
Q_u=\sum_{i\in\mathcal{I}}\sum_{j\in\mathcal{J}}\frac{\alpha_u^{i,j}\lambda_u}{x^{i,j}C}
\end{eqnarray}
where $x^{i,j}$ denotes the number of computing RBs allocated to order index $j$ of NOMA cluster $i$.
\section{Optimization Problem}\label{sec:optimization}
In this section, we formulate an optimization problem to minimize the summation of the energy consumption of all the users with constraint on the total transmission and computing time. In particular, we enforce a deadline $D_u$ as an upper limit on the total time of task $u$ as follows
\begin{eqnarray}\label{equ6}
T_u+Q_u\leq D_u
\end{eqnarray}
Constraint~(\ref{equ6}) is equivalent to the following data rate requirement
\begin{eqnarray}\label{equ7}
R_u\geq\frac{L_u}{D_u-Q_u}
\end{eqnarray}
where $Q_u\leq D_u$. Therefore, we propose to solve the following optimization problem,
\begin{eqnarray}\label{equ8}
\text{P1}:\underset{\alpha_u^{i,j},~\beta^{r,i}~p_u^r,~x^{i,j}}{\text{minimize}} \sum_{u\in\mathcal{U}}E_u\nonumber
\end{eqnarray}
\vspace{-.17in}
\begin{eqnarray}
s.t.~\text{C1}:R_u\geq\frac{L_u}{D_u-Q_u}~\forall u\in\mathcal{U}\nonumber
\end{eqnarray}
\vspace{-.17in}
\begin{eqnarray}
\text{C2}:Q_u\leq D_u~\forall u\in\mathcal{U}\nonumber
\end{eqnarray}
\vspace{-.17in}
\begin{eqnarray}
\text{C3}: \sum_{i\in\mathcal{I}}\sum_{j\in\mathcal{J}}\alpha_u^{i,j}=1~\forall u\in\mathcal{U}\nonumber
\end{eqnarray}
\vspace{-.17in}
\begin{eqnarray}
\text{C4}: \sum_{u\in\mathcal{U}}\sum_{j\in\mathcal{J}}\alpha_u^{i,j}\geq2~\forall i\in\mathcal{I}\nonumber
\end{eqnarray}
\vspace{-.17in}
\begin{eqnarray}
\text{C5}: \sum_{i\in\mathcal{I}}\beta^{r,i}\leq1~\forall r\in\mathcal{R}_f\nonumber
\end{eqnarray}
\vspace{-.17in}
\begin{eqnarray}
\text{C6}: \sum_{i\in\mathcal{I}}\sum_{j\in\mathcal{J}}x^{i,j}\leq M_c\nonumber
\end{eqnarray}
\vspace{-.17in}
\begin{eqnarray}
\text{C7}: \alpha_u^{i,j}\leq\alpha_u^{i,j-1}~\forall u\in\mathcal{U},~i\in\mathcal{I},~2\leq j\leq u_{max}\nonumber
\end{eqnarray}
\vspace{-.17in}
\begin{eqnarray}
\text{C8}:\sum_{r\in\mathcal{R}_f} p_u^r\leq P_u^{max}~\forall u\in\mathcal{U}\nonumber
\end{eqnarray}
\vspace{-.17in}
\begin{eqnarray}
\text{C9}: p_u^r\geq0~\forall u\in\mathcal{U},~r\in\mathcal{R}_f\nonumber
\end{eqnarray}
\vspace{-.17in}
\begin{eqnarray}
\text{C10}: \alpha_u^{i,j}\in\{0,1\}~\forall u\in\mathcal{U},~i\in\mathcal{I},~j\in\mathcal{J}\nonumber
\end{eqnarray}
\vspace{-.17in}
\begin{eqnarray}
\text{C11}: \beta^{r,i}\in\{0,1\}~\forall r\in\mathcal{R}_f,~i\in\mathcal{I}\nonumber
\end{eqnarray}
\vspace{-.17in}
\begin{eqnarray}
\text{C12}: x^{i,j}\in\mathds{N}~\forall i\in\mathcal{I},~j\in\mathcal{J}\nonumber
\end{eqnarray}
Inequality constraints C1 is the computing aware minimum data rate requirement per user. $Q_u$ is upper bounded by $D_u$ in constraint C2. The equality constraint C3 is to ensure that each user is assigned to only one NOMA cluster and also only one order index within that NOMA cluster. Constraint C4 is to ensure that at least two users are assigned to each cluster. In addition, we use the inequality constraints C5 to ensure that each frequency RB is allocated to only one cluster and C6 to bound the total allocated computing RBs by the available computing RBs. Constraint C7 is designed to give assignment priority to a lower value of order in one NOMA cluster over all the higher values of order in the same cluster. Moreover, by constraints C8, the total transmission power of user $u$ is limited to power budget $P^{max}_{u}$. Finally, constraint C9 restricts variable $p_u^r$ to positive values, constraints C10 and C11 restrict the variables $\alpha_u^{i,j}$ and $\beta^{r,i}$ to binary choices, and constraint C12 is to restrict variables $x^{i,j}$ to the integer values.

Note that P1 defines a flexible Mixed Integer Non-Linear Programming (MINLP) problem which involves binary, integer and real variables.
However, finding an optimal solution to this problem is intractable
and presents computing complexity where the complexity grows fast with the number of variables.
Furthermore, the objective function in P1 is not generally a convex function.
Therefore, in order to reduce the complexity and obtain high quality solutions in reasonable time, we follow a two-phase approach. First, we propose an efficient heuristic algorithm for user clustering and RBs allocation. In fact, the heuristic algorithm is designed to decide about the binary and the integer variables. Second, having the binary and integer variables removed, we formulate a convex optimization problem for the transmission power control to be solved independently per NOMA cluster.

\subsection{User Clustering and RBs allocation}\label{sec:heuristic}
The pseudo code for the user clustering and RBs allocation is summarized in Algorithms~\ref{alg:1}. As shown in this algorithm, we carry out the user clustering, computing RBs allocation and frequency RBs allocation in three separated phases.
Denote $\bar{h}_u=\frac{\sum_{r\in\mathcal{R}_f}h_u^r}{M_f}$ as the average channel condition of user $u$. In the clustering phase (lines 2-9), we follow a clustering method based on the average channel conditions.
In other words, the users with the higher average channel gains are assigned to the lower order indices of the clusters.
By doing so, a user with a higher channel gain does not interfere to the users with the lower channel gains since its interference is canceled out by the SIC receiver. Thus, the users with higher channel gains can transmit with the maximum transmission power, thereby, improving the sum-rate of the cluster.
Let $u^{i,j}$ be the user assigned to the $j$th index of cluster $i$ and $\mathcal{U}^i$ as the set of all the users assigned to cluster $i$.
Note that $u^{i,j}$ and $\mathcal{U}^i$ are known after the clustering phase.
In the next phase, i.e., the computing RBs allocation phase (lines 11-31), we first allocate the computing RBs to satisfy condition $Q_u<D_u$ for all the users. Here, we assume the available number of computing RBs are sufficient to satisfy such a condition. Then, for each of the remaining computing RBs, we search the set of all the clusters to find a favorite cluster, i.e., $\hat{i}$.
Denote $Q_{u^{i,j}}^{x^{i,j}+1}$ as the computing time of user $u^{i,j}$ if we increase $x^{i,j}$ to $x^{i,j}+1$.
The favorite user is identified by comparing the terms $\frac{L_{u^{i,j}}}{D_{u^{i,j}}-Q_{u^{i,j}}^{x^{i,j}}}-\frac{L_{u^{i,j}}}{D_{u^{i,j}}-Q_{u^{i,j}}^{x^{i,j}+1}}$. By doing so, we not only take into consideration of the input size of the users but also their deadlines in the computing resource allocation.

The procedure for the frequency RBs allocation phase is presented in lines 33-54. As shown in lines 33-43, we initially allocate the channels to satisfy the minimum data rate requirement for all the users. Denote $\mathcal{R}^i$ as the set of channels allocated to cluster $i$ and $\mathcal{I}'$ as the set of clusters consisting of some users with unsatisfied minimum data rate requirement. Let $\mathcal{R}_{f_0}$ be the set of already allocated channels. The favorite cluster for each channel, i.e., $\hat{i}$, is the cluster that achieves the best sum-rate
over that channel as compared to those clusters which have users with unsatisfied minimum data rate. The sets $\mathcal{R}_{f_0}$, $\mathcal{R}^i$, and powers $P_u^r$ are accordingly updated after each channel allocation. Note that we assume the maximum transmit power of each user is equally divided among all the allocated channels to its corresponding cluster.
Then, for each of the remaining frequency RBs to be allocated, we search over all the clusters and identify a favorite cluster by comparing $\sum_{u\in\mathcal{U}_i}(\frac{L_u}{R_u^{\mathcal{R}^i}}-\frac{L_u}{R_u^{\mathcal{R}^i\cup r}})P_u^{max}$ (lines 44-54). Here,
$R_u^{\mathcal{R}^i}$ and $R_u^{\mathcal{R}^{i\cup r}}$ denote the data rate of user $u$ based on the current allocation and that of based on the current allocation as well as the allocation of RB $r$ to cluster $i$, respectively.
Note that these data rates are calculated based on the current values of $p_u^r$.
In fact, the favorite cluster for each channel is the one that achieves the maximum increase in the objective function of~P1. After identification of favorite cluster $\hat{i}$, we accordingly update set $\mathcal{R}^{\hat{i}}$ as well as powers $P_u^r$ for all the users in $\mathcal{U}^{\hat{i}}$.

\begin{algorithm}
\caption{}
\label{alg:1}
\begin{algorithmic}[1]
\STATE \textbf{User Clustering}
\STATE sort the users in $\mathcal{U}$ such that $\bar{h}_1\geq \bar{h}_2\geq...\geq \bar{h}_U$
\FORALL {$j\in\mathcal{J}$}
\IF {$jN\leq U$}
\STATE assign users $\{(j-1)N+1,..., jN\}$ as the $j$-th users of clusters $\{1,...,N\}$, respectively
\ELSE
\STATE assign users $\{(j-1)N+1,..., U\}$ as the $j$-th users of clusters $\{1,...,U-(j-1)N\}$, respectively
\ENDIF
\ENDFOR
\STATE \textbf{Computing RBs Allocation}
\FORALL {$i\in\mathcal{I}$}
\FORALL {$j\in\mathcal{J}$}
\REPEAT
\STATE {$x^{i,j}=x^{i,j}+1$}
\STATE {$M_c=M_c-1$}
\UNTIL condition $Q_{u^{i,j}}<D_{u^{i,j}}$ is satisfied
\ENDFOR
\ENDFOR
\REPEAT
\STATE $\hat{q}=0$, $\hat{i}\gets\emptyset$ and $\hat{j}\gets\emptyset$
\FORALL {$i\in\mathcal{I}$}
\FORALL {$j\in\mathcal{J}$}
\IF {$\frac{L_{u^{i,j}}}{D_{u^{i,j}}-Q_{u^{i,j}}^{x^{i,j}}}-\frac{L_{u^{i,j}}}{D_{u^{i,j}}-Q_{u^{i,j}}^{x^{i,j}+1}}\geq\hat{q}$}
\STATE $\hat{i}\gets i$ and $\hat{j}\gets j$
\STATE $\hat{q}\gets \frac{L_{u^{i,j}}}{D_{u^{i,j}}-Q_{u^{i,j}}^{x^{i,j}}}-\frac{L_{u^{i,j}}}{D_{u^{i,j}}-Q_{u^{i,j}}^{x^{i,j}+1}}$
\ENDIF
\ENDFOR
\ENDFOR
\STATE $x^{\hat{i},\hat{j}}=x^{\hat{i},\hat{j}}+1$
\STATE $M_c=M_c-1$
\UNTIL $M_c>0$
\STATE \textbf{Frequency RBs Allocation}
\STATE $p_u^r=P_u^{max}~\forall u\in\mathcal{U},r\in\mathcal{R}_f$
\STATE {$\mathcal{R}_{f_0}\gets\emptyset$, $\mathcal{I}'\gets\mathcal{I}$ and $\mathcal{R}^i\gets\emptyset$}
\FORALL {$r\in\mathcal{R}_f$}
\STATE $\hat{i}=\underset {i\in\mathcal{I}'}{arg max}~(\sum_{u\in\mathcal{U}^i}R_u)$
\STATE $\beta^{r,\hat{i}}=1$, $\mathcal{R}_{f_0}\gets\mathcal{R}_{f_0}\cup r$ and $\mathcal{R}^{\hat{i}}=\mathcal{R}^{\hat{i}}\cup r$
\STATE $P_u^r=\frac{P_u^r}{|\mathcal{R}^i|+1}~\forall u\in\mathcal{U}^{\hat{i}},r\in\mathcal{R}_f$
\IF {$R_u\geq\frac{L_u}{D_u-Q_u}~\forall u\in\mathcal{U}^{\hat{i}}$}
\STATE {$\mathcal{I}'\gets\mathcal{I}'\setminus\hat{i}$}
\ENDIF
\ENDFOR
\STATE $\mathcal{R}_f=\mathcal{R}_f\setminus\mathcal{R}_{f_0}$
\FORALL {$r\in\mathcal{R}_f$}
\STATE $\hat{e}=0$ and $\hat{i}\gets\emptyset$
\FORALL {$i\in\mathcal{I}$}
\IF {$\sum_{u\in\mathcal{U}^i}(\frac{L_u}{R_u^{\mathcal{R}^i}}-\frac{L_u}{R_u^{\mathcal{R}^i\cup r}})P_u^{max}\geq\hat{e}$}
\STATE $\hat{i}\gets i$
\STATE $\hat{e}\gets\sum_{u\in\mathcal{U}_i}(\frac{L_u}{R_u^{\mathcal{R}^i}}-\frac{L_u}{R_u^{\mathcal{R}^i\cup r}})P_u^{max}$
\ENDIF
\ENDFOR
\STATE $\beta^{r,\hat{i}}=1$ and $\mathcal{R}^{\hat{i}}\gets\mathcal{R}^{\hat{i}}\cup r$
\STATE $P_u^r=\frac{P_u^r}{|\mathcal{R}^i|+1}~\forall u\in\mathcal{U}^{\hat{i}},r\in\mathcal{R}_f$
\ENDFOR
\end{algorithmic}
\end{algorithm}

In problem P1, we do not take into consideration of contiguous RB constraint. Under RB contiguity constraint, all the RBs allocated
to a single user must be contiguous in frequency~\cite{lee2009proportional}. Therefore, one can extend problem P1 by adding the following contiguity constraint,
\begin{eqnarray}\label{equ33}
\sum_{r=k_1}^{k_2}\beta^{r,i}=k_2-k_1+1~\forall i\in\mathcal{I},\beta^{k_1,i}=\beta^{k_2,i}=1
\end{eqnarray}
While the user clustering and computing RBs allocation phases in Algorithm~\ref{alg:1} are still valid to solve the extended problem under constraint~(\ref{equ33}), the frequency RBs allocation must be changed to comply with this constraint. Notice that the frequency RBs allocation under the contiguity constraint is known to be NP-hard~\cite{lee2009proportional}. To address this hardness, we can adopt the RB grouping algorithm proposed in~\cite{lee2009proportional}.
In a nutshell, $M_f$ available frequency RBs are divided into $n$ groups, and the frequency RBs allocation phase in Algorithm~\ref{alg:1} is accomplished with the granularity of RB groups. As a result, by
extending the unit of consideration from a single RB to the RB group, there is a wider view to obtain optimal frequency RB allocation.
\subsection{Power Control}
Given the user clustering and frequency RBs allocation, the binary variables $\alpha_u^{i,j}$ and $\beta^{r,i}$ in problem P1 are fixed to 0 or 1. Given the computing RBs allocation, the integer constraints vanish and the power control per NOMA cluster follows.
In fact, we can now eliminate all the terms that do not depend on the transmit
powers.
However, the objective function of problem P1 is non-convex in transmit powers. To this end, for the power control, we propose to minimize the total power consumption of the users instead of their energy consumption since the communication time is taken care of in constraint C1.
In other words, power consumption minimization problem can provide an accurate solution for the energy consumption minimization as the communication time is upper bounded in constraint C1.
Let consider $i$ as an NOMA cluster of interest. Denote $\mathcal{U}^i=\{1,...,u_{max}\}$ as the set of the users assigned to order indices $\{1,...,u_{max}\}$ of cluster $i$, respectively. Recall that $\mathcal{R}^i$ is the set of frequency RBs allocated to cluster $i$. Therefore, the power control optimization problem for cluster $i$ can be written as,
\begin{eqnarray}\label{equ9}
\text{P2}:\underset{p_u^r}{\text{minimize}}\sum_{u\in\mathcal{U}^i}
\sum_{r\in\mathcal{R}^i}p_u^r \nonumber
\end{eqnarray}
\vspace{-.15in}
\begin{eqnarray}
s.t.~\text{C1}:R_u\geq\frac{L_u}{D_u-Q_u}~\forall u\in\mathcal{U}^i\nonumber
\end{eqnarray}
\vspace{-.15in}
\begin{eqnarray}
\text{C2}:\sum_{r\in\mathcal{R}^i} p_u^r\leq P_u^{max}~\forall u\in\mathcal{U}^i\nonumber
\end{eqnarray}
\vspace{-.17in}
\begin{eqnarray}
\text{C3}: p_u^r\geq0~\forall u\in\mathcal{U}^i,~r\in\mathcal{R}^i\nonumber
\end{eqnarray}
Nevertheless, we can prove the following theorems:
\begin{theorem}
Optimization problem P2 is equivalent to the optimization problem P3.
\end{theorem}
\begin{eqnarray}\label{equ10}
\text{P3}:\underset{S_u^r,Z_u^r,R_u}{\text{minimize}} \sum_{u\in\mathcal{U}^i}\sum_{r\in\mathcal{R}^i}e^{S_u^r} \nonumber
\end{eqnarray}
\vspace{-.15in}
\begin{eqnarray}
\text{C1}:R_u\geq\frac{L_u}{D_u-Q_u}~\forall u\in\mathcal{R}^i\nonumber
\end{eqnarray}
\vspace{-.15in}
\begin{eqnarray}
\text{C2}: \sum_{r\in\mathcal{R}^i}e^{S_u^r}\leq P_u^{max}~\forall u\in\mathcal{U}^i\nonumber
\end{eqnarray}
\vspace{-.15in}
\begin{eqnarray}
\text{C3}:Z_u^r\leq B\log(1+\frac{h_u^re^{S_u^r}}{\sigma^2+\sum_{k=u+1}^{u_{max}}h_k^re^{S_k^r}})~\forall u\in\mathcal{U}^i\nonumber
\end{eqnarray}
\vspace{-.15in}
\begin{eqnarray}
\text{C4}:R_u=\sum_{r\in\mathcal{R}^i}Z_u^r~\forall u\in\mathcal{U}^i\nonumber
\end{eqnarray}
\vspace{-.15in}
\begin{IEEEproof}
Problem P2 is not a convex problem in the current form. However, let consider the following change of variables
\begin{eqnarray}\label{equ11}
p_u^r=e^{S_u^r}~\forall u\in\mathcal{U}^i
\end{eqnarray}
where obviously~(\ref{equ11}) is an one to one mapping, and it is always possible to determine the variable
$S_u^r$ from $p_u^r$ and viceversa. We also add the new variables $Z_u^r$ and $R_u$ to the problem formulation, where $Z_u^r$ is given by
\begin{eqnarray}\label{equ12}
Z_u^r=B\log(1+\frac{h_u^re^{S_u^r}}{\sigma^2+\sum_{k=u+1}^{u_{max}}h_k^re^{S_k^r}})
\end{eqnarray}
and
\begin{eqnarray}\label{equ13}
R_u=\sum_{r\in\mathcal{R}^i}Z_u^r
\end{eqnarray}
Consequently, by substituting the new variables $S_u^r$, $Z_u^r$ and $R_u$ in P2, and after simple algebraic manipulation, we can obtain the formulation of problem P3. Note that in the formulation of P3, we have changed the equality relations $Z_u^r=B\log(1+\frac{h_u^re^{S_u^r}}{\sigma^2+\sum_{k=u+1}^{u_{max}}h_k^re^{S_k^r}})$ to the inequality constraints C3. This change is needed for the convexity of problem P3, and does not affect the optimal solution since the data rate of user $u$ on RB $r$ at the optimality cannot be less than $B\log(1+\frac{h_u^re^{S_u^r}}{\sigma^2+\sum_{k=u+1}^{u_{max}}h_k^re^{S_k^r}})$.
\end{IEEEproof}

\begin{theorem}
Optimization problem P3 is convex in the high SINR regime.
\end{theorem}
\begin{IEEEproof}
The objective function of P3 is the sum of exponentials and is thus convex. While it is straight forward to prove the convexity of constraints C1, C2, and C4, the inequality constraint C3 is not convex since the throughput function is a non-convex function of the powers. A commonly used solution to deal with the non-convexity of the throughput function is the approximation $\log(1+x)\approx\log(x)$ which is a valid approximation in the high SINR regime~\cite{qiu1999performance,julian2002qos,ram2009distributed}.
As a result of this approximation, constraint C3 becomes
\begin{eqnarray}\label{equ13}
Z_u^r+\log(\sigma^2h_u^{r^{-1}}e^{-S_u^r}+\sum_{k=u+1}^{u_{max}}h_u^{r^{-1}}h_k^re^{S_k^r-S_u^r})\leq0~\forall u\in\mathcal{U}^i\nonumber
\end{eqnarray}
where $\log(\sigma^2h_u^{r^{-1}}e^{-S_u^r}+\sum_{k=u+1}^{u_{max}}h_u^{r^{-1}}h_k^re^{S_k^r-S_u^r})$ is the log of a sum of exponentials, and thus is convex. Therefore, problem P3 is convex in the high SINR regime and the proof is complete.
\end{IEEEproof}

Nonetheless, problem P3 is convex and can be solved by efficient optimization techniques such as interior point methods. Note that the complexity of this problem may increase as the numbers of users per NOMA cluster increases. However, each NOMA cluster is assumed to be limited to a few number of users. Moreover, we assume that the channel coefficients $h_u^r$ are known only to the BS and this problem is solved centrally at the BS for each NOMA cluster.

\section{Simulation Results}\label{sim:results}
\begin{figure}
\center
\epsfig{file=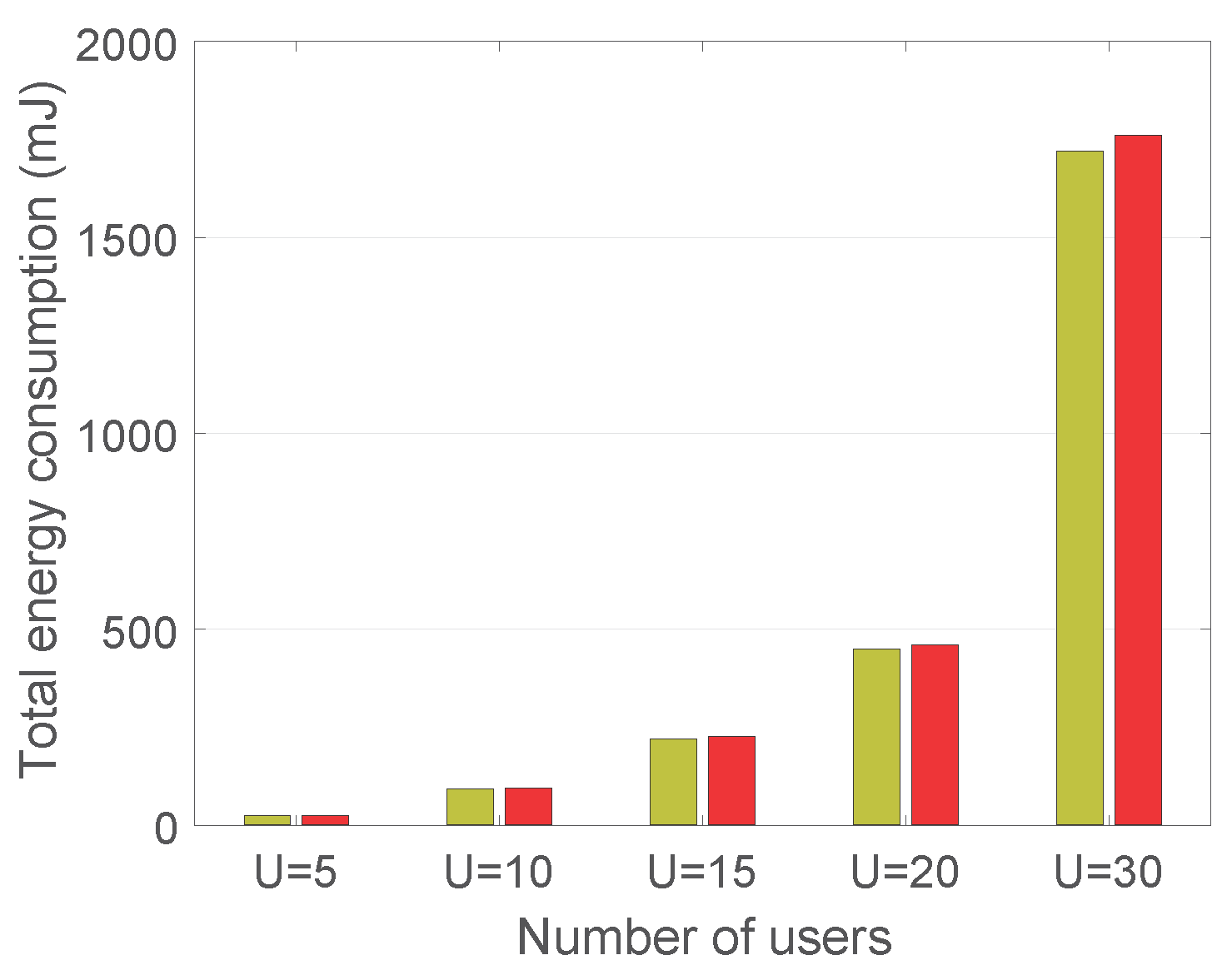,width=.6\linewidth,clip=}
\caption{Energy consumption comparison between heuristic and optimal approaches.}
\label{fig:1}
\end{figure}
\begin{figure}
\center
\epsfig{file=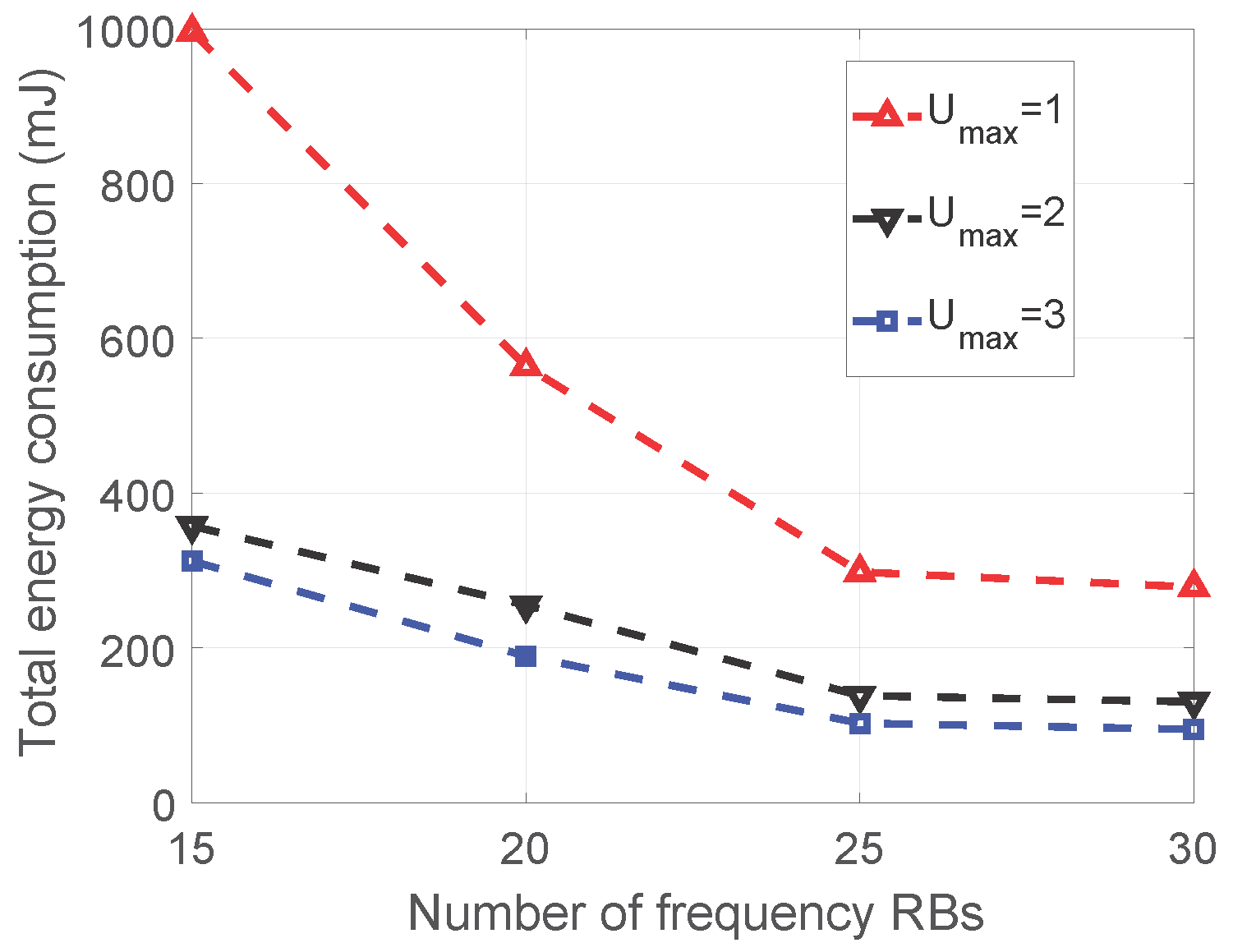,width=.6\linewidth,clip=}
\caption{Energy consumption of 10 users versus number of frequency RBs for $u_{max}$=1, 2 , 3, and 4.}
\label{fig:2}
\end{figure}
\begin{figure}
\center
\epsfig{file=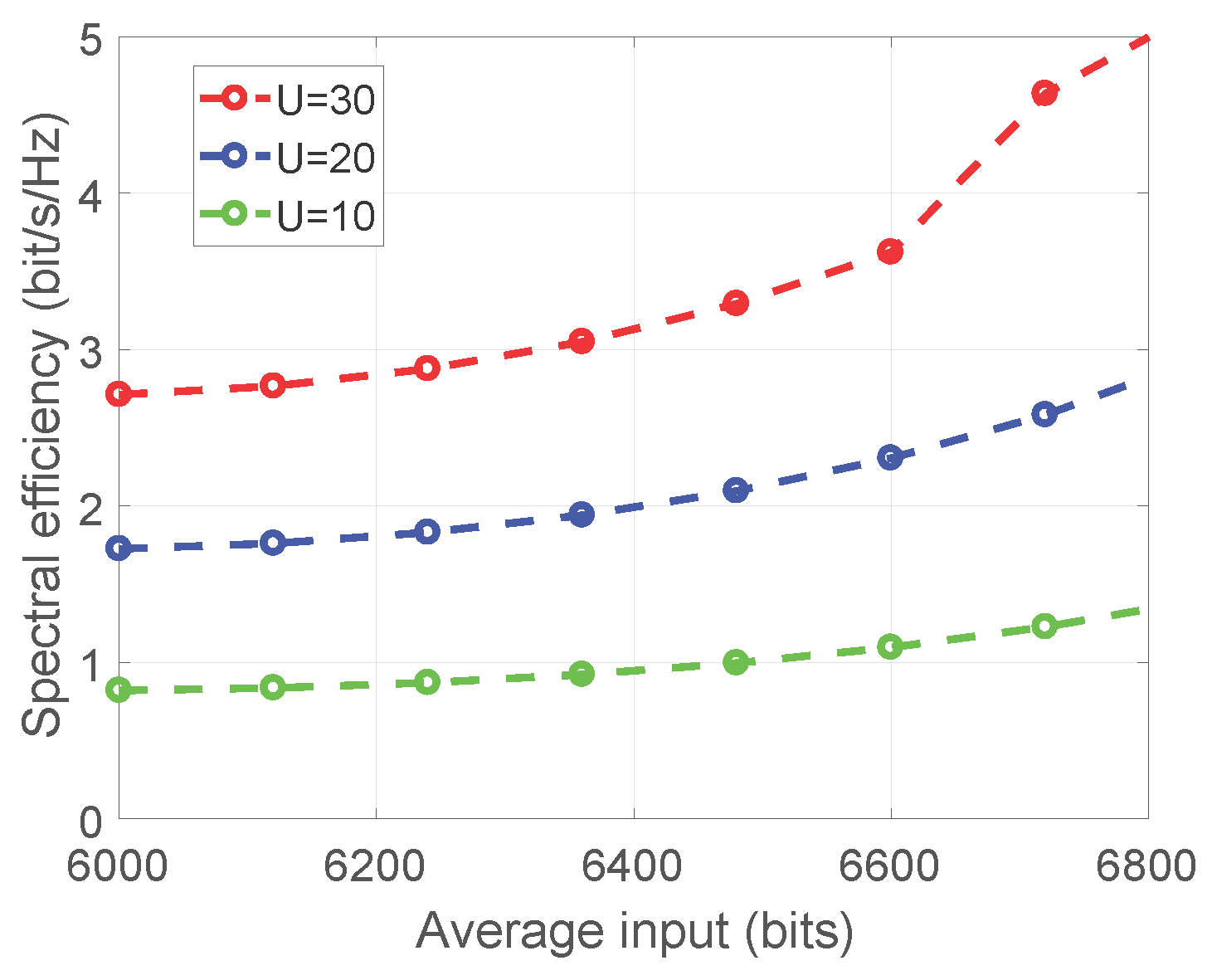,width=.6\linewidth,clip=}
\caption{Spectral efficiency versus average input.}
\label{fig:3}
\end{figure}
\begin{figure}
\center
\epsfig{file=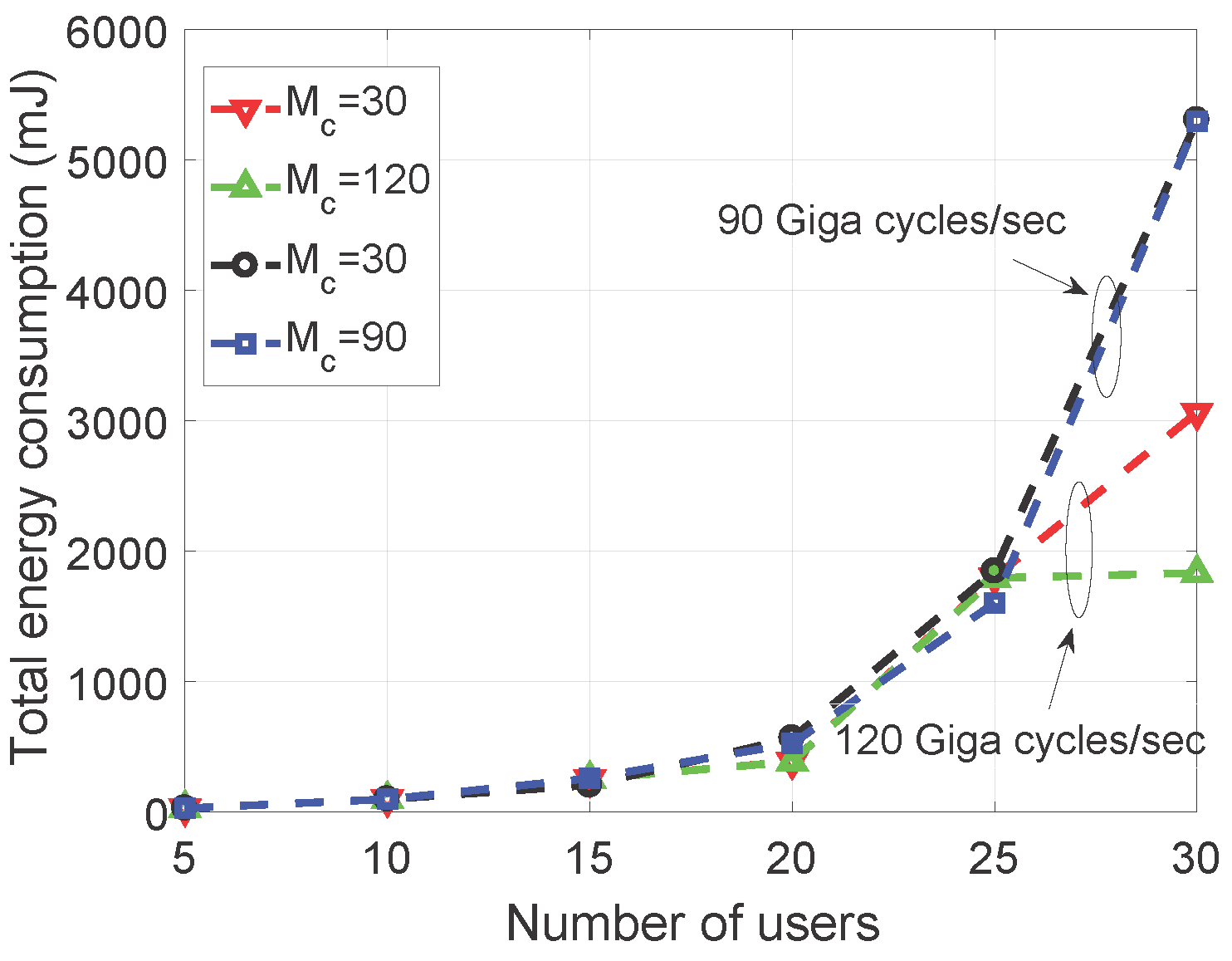,width=.6\linewidth,clip=}
\caption{Computing time versus number of users for different numbers of computing RBs and computing capacities.}
\label{fig:4}
\end{figure}
\begin{figure}
\center
\epsfig{file=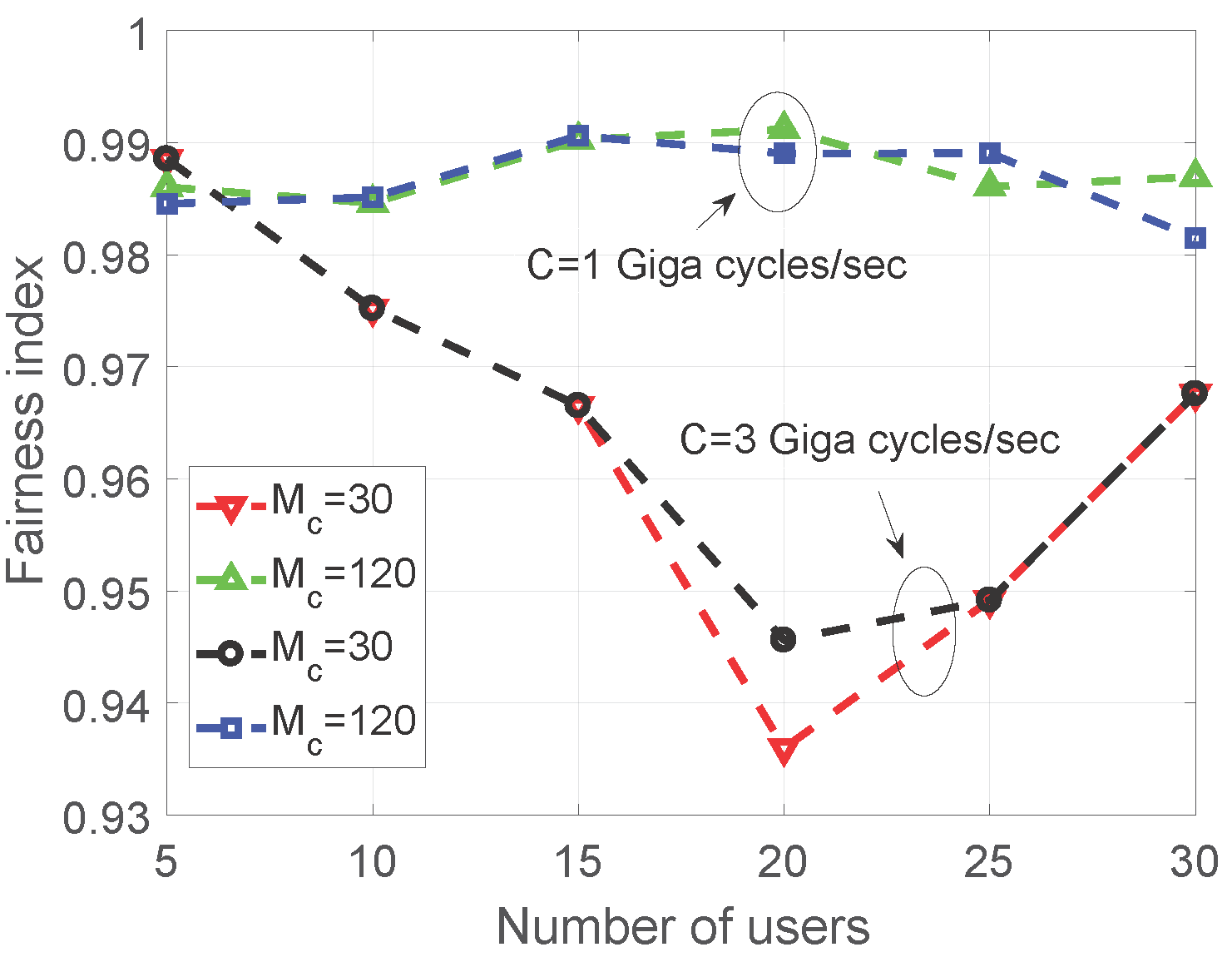,width=.6\linewidth,clip=}
\caption{Computing time fairness index versus number of users for different numbers of computing RBs and computing capacities.}
\label{fig:5}
\end{figure}
In this section, we evaluate the performance of the proposed edge computing aware NOMA scheme. We consider a single-cell with
1 km radius in which the users are uniformly distributed within the cell. We set the maximum transmit power of each user to 1 W, the maximum available frequency RBs to 30, and the bandwidth of each resource block to 180 kHz. The ITU pedestrian B fast fading model, the COST231 Hata propagation model for micro cell environment~\cite{al2014uplink,3GPP}, and the Lognormal  shadowing with 8 dB standard deviation
are implemented. The noise power spectral density is set to be 173 dBm/Hz.
Unless it is stated otherwise, we assume a total computing capacity of 300 Giga cycles per second is available as 30 computing RBs each with a capacity of 10 Giga cycles per second. The workload of each user ($\lambda_u$) is randomly generated according to a uniform distribution between 0.5 Giga cycles and 1 Giga cycles. The input of each user ($L_u$) and the deadline ($D_u$) are also randomly generated according to uniform distributions between 5000 bits and 7000 bits, and 400 ms to 500 ms, respectively.

Fig.~\ref{fig:1} compares the performance of the heuristic (Algorithm~\ref{alg:1}) and optimal (problem P1) approaches by providing the total energy consumption for different number of users where $u_{max}=3$. As we can see in this figure, the heuristic algorithm incurs a total energy consumption quite close to that of the optimal approach.
Meanwhile, the heuristic algorithm provides the suboptimal solution within a few seconds while the computation time of the optimal approach grows fast with the number of users.

Fig.~\ref{fig:2} illustrates the effect of $u_{max}$, i.e., maximum number of users allowed to share a frequency RB, on the total energy consumption of 10 users for different number of available frequency RBs. As shown in this figure, first, the energy consumption decreases by increasing the available number of frequency RBs; second, the energy consumption improves by increasing $u_{max}$ from 1 to 2, and 3 due to the fact that the spectral efficiency improves by increasing $u_{max}$. The energy consumption gap is more significant between $u_{max}=1$ and $u_{max}=2$ or 3 as compared to that of between $u_{max}=2$ and $u_{max}=3$. This result is attributed to the fact that the intra-cell interference becomes more considerable by increasing $u_{max}$. Fig.~\ref{fig:3} is also provided to evaluate the performance of the proposed scheme with $u_{max}=3$ in terms of the spectral efficiency. As the results show, the spectral efficiency is increased by increasing both the number of users and the average input of each user.

Fig~\ref{fig:4} shows the impact of the total computing capacity as well as its division into the computing RBs on the total energy consumption. In particular, we assume two different cases in terms of the total computing capacity, each with two different division scenarios. In the first case, we consider a total computing capacity of 120 Giga cycles per second, which is divided into 30 computing RBs, each with a capacity of 3 Giga cycles per second and 120 computing RBs each with a capacity of 1 Giga cycles per second as the first and second scenarios, respectively.
Similarly, for the second case, the total computing capacity of 90 Giga cycles per second is assumed to be divided to 30 RBs as the first scenario and 90 RBs as the second scenario. As Fig~\ref{fig:4} shows, the energy consumption can be improved specifically for higher number of users if the computing capacity is divided into smaller RBs, i.e., the scenarios with 120 and 90 RBs. This observation is due to the fact that the users have different amounts of workloads, and thus the computing capacity can be allocated more fairly to the users when it is divided to smaller blocks. Moreover, this improvement is more considerable for the case with the capacity of 120 Giga cycles per second since the difference between the two division scenarios is more pronounced. To understand the reason of the observation in Fig~\ref{fig:4}, we should analyze the performance of the computing RBs allocation scheme in terms of the fairness. To this end, we have adopted the Jain's fairness index~\cite{jain1984quantitative} for the computing time,
\begin{eqnarray}\label{equ14}
\text{Fairness index}=\frac{(\sum_{u=1}^{U}Q_u)^2}{U\sum_{u=1}^{U}(Q_u)^2}
\end{eqnarray}
Fig~\ref{fig:5} shows the Jain's fairness index which is bounded between of 0 and 1 for the aforementioned scenarios. As we can see in this figure, the scenarios with computing RBs with granularity of 1 Giga cycles per second are fairer as compared to those with granularity of 3 Giga cycles per second.

\section{conclusion}\label{conclude}
In this study, we have proposed an edge commuting aware NOMA technique, which can leverage the gains of uplink NOMA in reducing MEC users' energy consumption.
Specifically, we have formulated a NOMA based optimization framework that minimizes the energy consumption of MEC users via optimizing the user clustering, computing and communication resource allocation, and transmit powers. In particular, we have investigated the joint allocation of the frequency and computing RBs to the users that are assigned to different order indices within the NOMA clusters.
We have also designed an efficient heuristic algorithm for user clustering and RBs allocation, and formulated a convex optimization problem for the power control to be solved independently per NOMA cluster.
Moreover, we have evaluated and demonstrated the effectiveness of the proposed
NOMA scheme in lowering the energy consumption via simulations.

\bibliographystyle{IEEEtran}
\bibliography{ref}

\end{document}